# An expert consensus on identification likelihood in NeuroImaging data.

- Pernet, Cyril R. — Neurobiology Research Unit, Rigshospitalet & Department of Health Technology, DTU (Denmark) ORCID: 0000-0003-4010-4632
- Arvaneh, Mahnaz — University of Sheffield, Dept. of Electronic & Electrical Engineering (UK) ORCID: 0000-0002-5124-3497
- Bannier, Elise — CHU Rennes; Empenn team (Inria/INSERM/IRISA), Rennes (France) ORCID: 0000-0002-8942-7486
- Breakspear, Michael — University of Newcastle & Hunter Medical Research Institute (Australia) ORCID: 0000-0003-4943-3969
- Calhoun, Vince — Tri-Institutional Center for Translational Research in Neuroimaging and Data Science (TReNDS), Georgia State, Georgia Tech, Emory, Atlanta (USA), ORCID: 0000-0001-9058-0747
- da Silva Castanheira Jason - University College London, United Kingdom ORCID: 0000-0002-7022-9009
- Delorme, Arnaud — Swartz Center for Computational Neuroscience, University of California, San Diego (USA) ORCID: 0000-0002-0799-3557
- Eickhoff, Simon — Forschungszentrum Jülich, INM-7 & Heinrich Heine University Düsseldorf, Institute of Systems Neuroscience (Germany) ORCID: 0000-0001-6363-2759
- Eklund Anders -- Linköping University, Department of Biomedical Engineering and Department of Computer and Information Science (Sweden), ORCID: 0000-0001-7061-7995
- Fornito, Alex — Turner Institute for Brain & Mental Health, School of Psychological Sciences, and Monash Biomedical Imaging, Monash University, Melbourne, Australia. ORCID: 0000-0001-9134-480X
- Ganz, Melanie — University of Copenhagen (DIKU) & Neurobiology Research Unit, Rigshospitalet (Denmark), ORCID:0000-0002-9120-8098
- Moreau, Clara — Université de Montréal & CHU Sainte-Justine Research Center (Canada), ORCID: 0000-0001-6217-731X
- Nichols, Tom — University of Oxford, Big Data Institute (UK) ORCID: 0000-0002-4516-5103
- Nilsonne Gustav — Karolinska Institutet (Sweden) & Swedish National Data Service (Sweden), ORCID: 0000-0001-5273-0150
- Oostenveld, Robert — Donders Institute, Radboud University (Netherlands) ORCID: 0000-0002-1974-1293
- Paton, Bryan, University of Newcastle (Australia) ORCID: 0000-0001-6777-5001
- Ritter, Petra — Charité hospital, Universitätsmedizin Berlin, Brain Simulation Section (Germany) ORCID: 0000-0002-4643-4782
- Rokem, Ariel — Krembil Centre for Neuroinformatics, Campbell Family Mental Health Research Institute, Centre for Addiction and Mental Health & Department of Psychiatry, University of Toronto (Canada),ORCID: 0000-0003-0679-1985
- Schirner Michael — Charité hospital, Universitätsmedizin Berlin, Brain Simulation Section (Germany), ORCID: 0000-0001-8227-8476
- Shine, James Mac - University of Sydney, (Australia) ORCID: 0000-0003-1762-5499

- Thiebaut de Schotten, Michel — CNRS, Institut des Maladies Neurodégénératives / Bordeaux Neurocampus (France) ORCID: 0000-0002-0329-1814
- Thirion, Bertrand — Inria , CEA, Université Paris-Saclay, ORCID: 0000-0001-5018-7895
- Veronese, Mattia — Padova University, Department of Information Engineering (Ita); King's College London, Dept. of Neuroimaging (UK)  ORCID: 0000-0003-3562-0683
- Voets, Natalie, Nuffield Dept. of Clinical Neurosciences, University of Oxford, UK ORCID: 0000-0001-8078-4471
- Yeo, B.T. Thomas — National University of Singapore, Centre for Sleep & Cognition, Yong Loo Lin School of Medicine  (Singapore) 0000-0002-0119-3276
- Jwa, Anita (Sue) — Stanford University, Department of Psychology (USA) ORCID: 0000-0002-1462-424X

# Abstract

Guidelines on the anonymity of neuroimaging data help manage research data, inform institutional legal and governance frameworks, and support data-sharing workflows. The question of identifiability is particularly salient in human brain imaging because such data are not only potentially personal (i.e., related to an identifiable person) but also sensitive and thus require extra precautionary and protective processes. The identifiability of brain imaging data is also salient for emerging legal and ethical frameworks regarding rights that emphasise the special status of brain data as closely linked to personal identity, mental privacy, and autonomy. Using a combined nominal group technique and Delphi process, we obtained expert consensus ratings of re-identification and singling-out likelihood across 34 neuroimaging data types, spanning multiple imaging modalities and processing levels, from curated research data to maps, matrices, and summary tables. Expert ratings suggest that 38–67% of the assessed neuroimaging data types could be shared with minimal restrictions based on their intrinsic re-identification likelihood, subject to contextual safeguards. The resulting classification enables researchers to engage with relevant legal teams to guide the appropriate and safe release of data, and to determine the appropriate level of protection (licensing, access controls, governance) in line with domestic and international legal requirements, with the ultimate goal of facilitating international data exchange. A website (https://brainimagingprivacy.github.io/) has been created to facilitate the dissemination of these findings and related evidence among relevant parties.

# Introduction

Sharing of research data has become an increasingly important scientific practice, motivated by its ability to improve transparency and reproducibility, enabling the aggregation of data across studies for large-scale and meta-analytic investigations, facilitating education and training, and allowing existing data to address questions beyond those envisioned by the original investigators (Poline et al., 2012). Data reuse can also substantially increase the return on research investment: analysing the International Neuroimaging Data-sharing Initiative (INDI), Milham et al. (2018) identified 913 publications based on shared data and estimated that generating the data used in these studies de novo would have cost approximately US$0.89–1.71 billion. Despite these benefits, data sharing remains poorly aligned with traditional academic incentives because researchers largely bear the effort and cost of acquiring, curating, and maintaining shareable data, while subsequent discoveries and academic credit may accrue to others. Consequently, adoption has historically depended largely on individual scientists' motivation, supplemented by funding-agency requirements.

A major countervailing concern to data sharing is protecting research participants. It is generally argued that even well-curated data, i.e. pseudonymized or de-identified data (see glossary), may allow individuals to be re-identified, linking their identity to sensitive information and creating informational or discriminatory harms. Structural Magnetic Resonance Imaging (MRI) provides a particularly concrete illustration of this problem because automated face-recognition methods can reconstruct facial surfaces from images and match them against photographs, for instance from social media. In a sample of 182 participants, Prakaashana et al. (2025) reported correct top-ranked identification of MRI-derived facial reconstructions in up to 59% of cases using freely available open-source software and 92–98% using commercial systems. This result should not, however, be interpreted directly as the probability of re-identification under realistic conditions since each MRI had a corresponding known photograph to be matched with. Modelling a substantially larger pool of potential photographs/identities, Jwa et al. (2024) showed that identification accuracy for defaced structural images decreased to 9.6% with 157 potential matches, 0.8% with 6,500 and 0.03% with 865,000 potential matches, suggesting that while re-identification is technically possible, its real-world likelihood (especially after defacing the MRI images) is considerably lower.

The EU General Data Protection Regulation (GDPR) defines ‘personal data’ as any information relating to an identified or identifiable natural person. An identifiable natural person is one who can be identified, directly or indirectly, in particular by reference to an identifier such as a name, an identification number, location data, an online identifier, or to one or more factors specific to the physical, physiological, genetic, mental, economic, cultural, or social identity of that natural person. GDPR excludes deceased persons’ data from its scope, although Member States may provide separate rules for their protection (Recital 27). While such a definition is not universally adopted, many other jurisdictions (e.g., Japan, the Republic of Korea, New Zealand, Argentina, Uruguay, the UK, and Switzerland) rely on closely related notions of identifiability to distinguish between data that can be shared openly (i.e., anonymised or de-identified data, see glossary) versus data that require protection. Given this definition and its broad use worldwide, it is necessary to identify which data are at risk of identification, meaning individuals can be recognised (e.g. directly identified from their name or picture) vs. ‘isolated’ (e.g. identified based on a combination of indirect but unique attributes). The European Data Protection Board (EDPB Guidelines 02/2026) also distinguishes record isolation, where a unique combination of attributes distinguishes a record within a dataset, from singling out, where that record can be associated with a particular individual using information and means reasonably likely to be available. Importantly, record isolation alone does not necessarily imply that an individual is identifiable. In the emerging context of neurorights (proposed protections for fundamental aspects of human autonomy and mental life, including cognitive liberty, mental privacy, mental integrity, and psychological continuity; Ienca & Andorno, 2017; Yuste et al., 2017), the ability of singling out is particularly salient. This is because advances in neuroimaging, machine learning, and large-scale data integration increasingly enable the recovery of stable, individual-specific patterns from data previously assumed to be anonymous. Tools that enable record isolation are useful technically and clinically. Technically, in a world where researchers use large databases and individuals travel and increasingly live in different cities and countries over their lives, it becomes useful to remove ‘duplicates’ (Puglisi et al., 2024) or leverage them to examine, for instance, changes over time. Clinically, record isolation means identifying unique patterns, which can serve as biomarkers or therapeutic targets. In our view, how much that ability to identify unique patterns from brain imaging enables singling out, i.e. re-identification, depends largely on how well those patterns link to public records. Proof-of-concept studies using internal databases (e.g., hospital records) show that singling out is possible, but such records require access control and are closely monitored (i.e. means are not reasonably likely). Technical capabilities therefore shape the boundary between personal and non-personal data as much as legal definitions do, and legal decisions should balance individual privacy risks against the scientific and societal benefits of acquiring, sharing, and reusing neuroscience data (Jwa & Poldrack, 2022).

This consensus supplies expert ratings differentiating re-identification from singling-out, here defined broadly. Ratings of ‘singling-out likelihood’ operationally reflect the ability of neuroimaging features to distinguish individuals, including through repeated measurements or matching across datasets (as such this is close to, but not equivalent to, the record isolation definition from the EDPB). Our ratings therefore characterise individual specificity and potential for linkage rather than determining identifiability, fitting with the broader international context and regulations. Re-identification risk requires a broader assessment considering e.g. contextual identification, including linkage with other data but also inference of individual attributes (e.g. EDPB Guidelines 02/2026, HIPAA 45 CFR, Australia Privacy Act/OAIC guidelines). For such inference assessment, security/privacy experts consider different types of attacks on data: background, linkage, or homogeneity. In a background attack, a third party has prior knowledge about a specific individual and can use it to identify that individual within a dataset. For example, if a neighbour knows a female participant took part in a study, they could re-identify her by looking at the participant's information data sheet leveraging certain types of known information such as age or occupation (Sweeney, 2002) and therefore establish the connection with her MRI data (even if defaced). In a linkage attack, third parties use external data sources to link anonymised data back to an individual (Narayanan & Shmatikov, 2008). For example, if a neuroimaging dataset contains overlapping information (such as age and postal code) with another publicly released dataset (e.g. a voter registration records), it may be possible to link the two datasets and re-identify individuals. Finally, in a homogeneity attack, a third party exploits a dataset's lack of diversity. If a particular characteristic (such as a medical condition or demographic detail) is the same for a group of people, information about the entire group can be inferred (Machanavajjhala et al., 2007). For example, if one knows a person participated in a neuroimaging study and the dataset shows that all participants share the same diagnosis, this homogeneity divulges that person's medical condition. This is of course also the case when one shares data from a single patient with anatomical or pathological characteristics which are particularly distinguishing, such as a rare disease (Boseley, 2016). Such inference assessment is beyond the scope of this work, but we encourage researchers and legal representatives to consider those cases, especially because neuroimaging data are usually shared along with demographic (age, sex, social economic status, etc.) and behavioural or clinical data. A recent analysis of participants' information (demographics and clinical scores, e.g. number from questionnaires and scales) from neuroimaging datasets found that clinical scores have a low overall likelihood of identification, but demographic data have a higher likelihood because many samples are likely of being unique (Kibsgaard et al., 2025) and inference from linkage using demographic and geolocalizing information has a high likelihood (Rocher et al., 2019).

In the following group consensus, we assessed the likelihood of identifying ‘normal’ brains across different brain imaging modalities and types (raw and derived brain scan data). Here raw data, means already curated data as per e.g. the Brain Imaging Data Standard (Gorgolewski et al., 2016; Poldrack et al., 2024). Additionally, normal brain means excluding brains with any visual abnormalities such as normal deviations in healthy volunteers (e.g. callosal agenesis), but it does include brains from common psychiatric diseases (they typically do not have visually identifiable features). We examine sharing data from patients with visible abnormalities (bones or brain) only in the discussion. Here, we describe how expert ratings were obtained and give our assessment of 34 imaging data types providing concrete, evidence-based guidance for researchers, IRBs, institutional offices, and policymakers in assessing reidentification likelihoods associated with those data.

# Method

The first author contacted the initial group of participants and invited participation in the neuroimaging data risk assessment, informing them of the overarching goal of the survey. Participants were also asked to invite other colleagues to collaborate. Although the group was formed ad hoc, we intentionally sought diversity in expertise across imaging modalities and in gender and geographical origin. Once the participant list was assembled, we used the nominal group and DELPHI methodologies synergistically (Fink et al., 1984) to reach consensus. Anyone who participated in the process (see the assessments and consensus below) gained co-authorship.

The expert assessment and consensus are limited to the most commonly used non-invasive structural or functional brain imaging methods in neuroscience research, where most data sharing occurs. This can be broadly subdivided by imaging modality: Magnetic Resonance

Imaging (MRI), Positron Emission Tomography (PET), Electroencephalography (EEG), Magnetoencephalography (MEG), and functional Near-Infrared Spectroscopy (fNIRS). Because TMS and tES brain-stimulation studies typically do not report directly on brain features, brain stimulation is not included in the current consensus.

## Item generation

Participants received an email to participate in this consensus (Annex 1). Participants were asked to generate a table (1) listing the different types of raw and derivative data they can think of, (2) rate them for risk of re-identification using a Likert scale from 1 to 5, and (3) indicate their self-reported level of expertise using a Likert scale from 1 to 5. They were also asked to add references to support their ratings.

To perform the rating, the email explicitly asked participants to classify data such that 5 is a high likelihood of re-identification (data can be seen as personal even after pseudonymization, they contain directly identifiable features, or features that can easily lead to re-identification either because of a low technical barrier or easy linkage with public sources such as social media, or open datasets of known individuals, e.g. Poldrack et al., 2015), 3 is medium risk (data have been pseudonymized, they contain unique features which could allow to single out individuals but such features either require expertise to extract or the linkage risk is deemed low) and 1 is no risk (data are fully anonymous and have no uniquely identifiable features). We recorded how many subjects generated the same item and computed the median and interquartile range (IQR) of ratings.

## Ratings

After initial individual-item generation and assessment, we constructed 34 data types (out of 51 items) and conducted a second evaluation, asking respondents to rate the likelihood of re-identification and of singling out individuals (Annex 2). We changed from a 1-5 scale to separate ratings of re-identification and singling out because response variance differed across data types during the item-generation round, suggesting one approach was used more than the other (see results). For this new rating, we formed two diverse expert groups based on self-reported expertise levels, and each completed a different survey. For both surveys, we organised items by data type (e.g., structural imaging, electrophysiology; see results), kept a fixed item order within each data type, and reversed the order of data types between the two groups to counteract order effects (Rasinski et al., 2012). The survey also provided supporting evidence for or against the likelihood of re-identification and of singling out individuals.

We first summarised expert ratings by computing the median and IQR. Second, we classified items into 4 categories based on their median rating: above or below 3. Finally, we computed Spearman correlations between re-identification and singling-out ratings for each category. All ratings and the analysis are available online at https://github.com/BrainImagingPrivacy/BrainImagingPrivacy.github.io/blob/main/papers/expert-ratings/ratings_analysis.ipynb.

## Consensus

Based on self-reported levels of expertise, three groups were formed to discuss results related to (1) structural imaging, diffusion imaging, and PET (7 respondents out of 10 invitees); (2) dynamic tomographic imaging, electrophysiology, and fNIRS (4 respondents out of 8 invitees); and (3) connectivity (5 respondents out of 9 invitees). Teams met online and were provided with the previous round's results to discuss ratings and evidence. They were instructed to 1st resolve 'undecided' items, then discuss linkage likelihood. This additional information is useful when considering items with a low likelihood of re-identification but a high likelihood of singling out. Since singling out is a precondition for re-identification, knowing whether linkage is possible helps. A brain imaging data type can have features unique to individuals but so abstract that no other data can be associated with it, thus rendering it low risk. Revised ratings were agreed through discussion within each subgroup.

Based on the subgroup ratings, each item was classified as belonging to one of three classes: (1) Identifiable, meaning that data contain directly identifiable features or features that can likely lead to re-identification; (2) Distinctive, meaning that the data pose no obvious re-identification risk but contain unique features that allow individuals to be singled out (such features, however, require expert domain knowledge for background attacks or entail high technological knowledge to be extracted and linked); and (3) Low likelihood, meaning that available evidence and expert ratings indicate limited likelihood for re-identification or singling out under the stated assumptions. Note that this rating based classification represents ordinal expert judgments; low ratings should therefore not be interpreted as evidence of zero risk. The entire group had access to these results during the article writing to resolve or clarify remaining issues with specific items.

# Results

## Items generation

We obtained 22 tables of generated items. These were analysed, homogenised, and assembled by the first author (CP) to allow extraction of the median rating scores. The results led to 7 data types: (1) structural data (MRI, CT, but also synthetic data), (2) Positron Emission Tomography, (3) Diffusion Imaging, (4) Dynamic imaging (functional MRI and ASL), (5) Magneto- and ElectroEncephalographic data, (6) fNIRs, and (7) connectivity. In total, the group had identified 51 brain imaging items. Synthetic data are in this context structural imaging data created using machine learning (ML) or generative AI models (Giuffrè & Shung, 2023; Koetzier et al., 2024). The models are trained on real data to learn the data distribution, and the outputs are images either from the same or different modality. A generative model can for example generate a T2w MRI image from a T1w image (Xiao et al., 2018), or a CT image from an T1w MRI image, and as such the identification likelihood is not higher or lower from these new images given that there exists a paired real image. Generative models can also generate completely new images from noise, and these images do not belong to a specific person. Generative AI models can in principle output an infinite number of new images, but the identification risk is unclear (Usman Akbar et al., 2025, Dar et al., 2025, Giuffrè & Shung, 2023; Koetzier et al., 2024).

Scoring of the items (annex 2) revealed large disparities across classes. Coupled with the raters' arguments and the cited literature, it is apparent that different classes, or even items within a class, were rated differently. Instructions asked to make a joint assessment of reidentification and singling out (annex 1), but results showed items were rated using either one or the other approach. For instance, raw structural images were mentioned in 20 tables out of 22 with a median rating of 5 and an interquartile range of 0, indicating strong agreement toward high identification likelihood (as per instruction, 5 means those data have a high likelihood of reidentification, defined as: 'data can be seen as personal even after pseudonymization, they contain directly identifiable features, or features that can easily lead to re-identification either because of a low technical barrier or easy linkage with public sources'). In contrast, task-based activation maps (fMRI) were obtained in 9 tables with a median of 1 but an interquartile range of 2.5. While most rated the likelihood of risk as low (defined as 'data are fully anonymous and have no uniquely identifiable features'), some raters noted that such data have uniquely discriminable features (not related to brain abnormalities as per instructions) and thus rated the risk much higher. This distinction within raters between re-identification and singling out depending on the item has then shaped the online ratings and consensus discussions.

# Ratings

Figure 1 shows the median and IQR for each data type re-organised into 34 items (individual values are available in the GitHub repository and Zenodo archive). The classification into 4 groups based on median ratings showed only one data type as high risk for re-identification and for being singled out: raw MRI and CT images. If an imaging data type is likely to lead to re-identification (identification risk > 3), individuals can also be singled out (single-out risk > 3). Therefore, one can expect no cases of simultaneous high identification risk and low single-out risk, which is what was observed in the ratings. However, we observed the reverse for 7 items: low re-identification likelihood and high singling-out likelihood. Notably, 13 data types received low ratings for both re-identification and singling out. The largest discrepancies were observed for ratings of the risk of being singled out, leaving 13 data types as 'undecided' (at least one median equal to 3). Those were the items discussed in priority by the online discussion groups: dMRI connectivity matrices, Event related electrophysiological response (ERP-ERF), Defaced MRI/CT data, Morphometric models, Raw DWI, Raw and Preprocessed electrophysiological data (EEG, MEG), Raw fMRI data, Raw or preprocessed fNIRS data, Space-Time decompositions from dynamic tomographic imaging (fMRI, fNIRS, PET), Spatially normalized dynamic data (fMRI, ASL), Synthetic structural MRI and CT - same image type, Table of morphometric (volume, thickness, etc) or qMRI (T1 time) information (ROI based), fNIRS connectivity matrices.

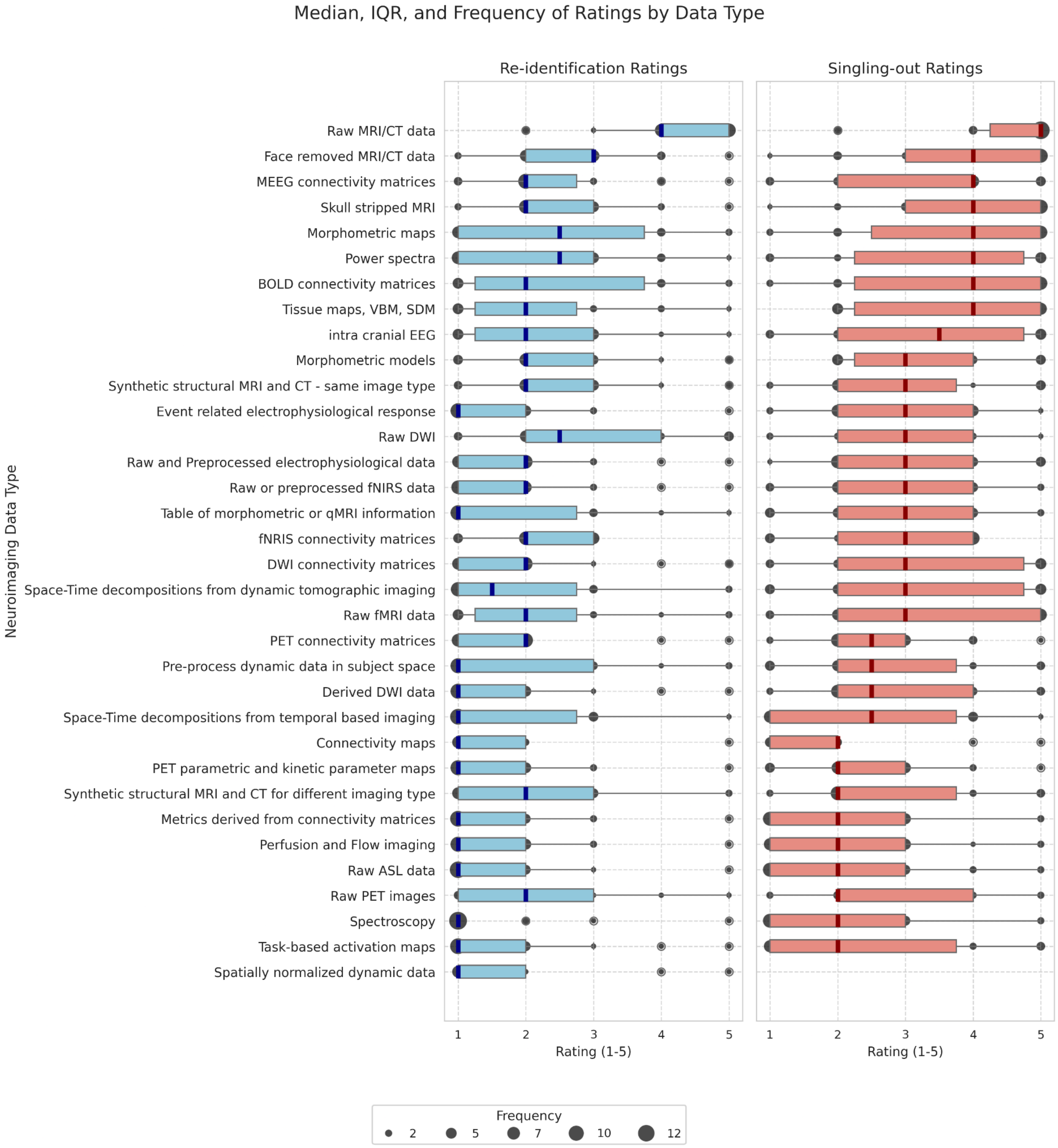

*Figure 1. Ratings for both re-identification (left, blue) and being singled out (right, red) risks. Black circles indicate the ratings given, and their size indicates how many raters gave that rating. Bars indicate the IQR, with the median shown as a thin bar in the middle. Items are sorted from the highest to the lowest single out ratings.*

The overall Spearman correlation (across all variables) between the rated risk of re-identification and the risk of singling out was 0.628 (p<0.001), ranging from 0.449 (fNIRS) to 0.68 (connectivity) across data type classes (Table 1).

| | Number of data points (number of raters * number of items) | Spearman correlation | FDR-corrected P-value |
|---|---|---|---|
| Structural Images | 162 | 0.562 | <0.001 |
| Dynamic Imaging (BOLD, ASL) | 126 | 0.656 | <0.001 |
| Diffusion Imaging | 27 | 0.502 | 0.011 |
| PET | 27 | 0.468 | 0.016 |
| Electrophysiology (MEEG) | 90 | 0.588 | <0.001 |
| fNIRS | 18 | 0.449 | 0.061 |
| Connectivity | 117 | 0.680 | <0.001 |

*Table 1: Spearman Correlation* between re-identification and singling-out ratings *for NeuroImaging* data type classes*. The number of data points reflects the number of items in each category multiplied by the number of ratings.*

# Consensus and recommendations

Based on the combination of re-identification and singling-out likelihoods consensus ratings, we distinguish data that are: (1) Identifiable; (2) Distinctive; and (3) Low likelihood (see method). The tables below present this classification together with the consensus ratings per data types identified during the item generation analysis. For items that remain undecided (i.e. with likelihoods of 3), diverse considerations about the data are also proposed. These should help judge, depending on context, whether the likelihood should be higher or lower than 3.

<u>Structural MRI/CT data (table 2):</u>

During group discussions, the expert subgroup changed the re-identification likelihood for defaced MRI/CT data from 3 to 2, considering the method and participant category (i.e., individuals without visible abnormalities). This matters because bone abnormalities can enable re-identification, and methods like blurring would be inadequate. Similarly, deblurring could reconstruct a face.

The raw diffusion-weighted imaging single-out rating also moved from 3 to 2, considering typical diffusion sequences in terms of resolution and number of directions. While sharing images labelled *_dwi (Gorgolewski et al., 2016) encompasses any type of diffusion imaging, we here distinguish basic 3-directional clinical DWI from more advanced diffusion MRI (dMRI). The latter, which is more typical in research settings, includes high resolution, high number of directions, and potentially multishell acquisition sequences allowing for reconstruction of fiber tracts, connectivity matrices and/or quantification of tissue metrics. The former refers to typical clinical acquisitions (single shell, few directions), typically generating only a summary apparent diffusion coefficient map. dMRI received a single-out likelihood rating of 4. Consistent with this distinction, dMRI connectivity matrices also received single-out ratings up to 4 (independently, since this was not the same expert subgroup; see method).

| Brain Imaging type | Evidence | Ratings and considerations |
|---|---|---|
| Raw structural MRI<br>Raw structural CT<br>- IDENTIFIABLE - | If coverage is extensive enough, it can be matched to photographs with regular face recognition software (Schwarz et al., 2019; Steeg et al., 2024) | ReID: 4.0 SingleOut: 5.0<br>Assuming research data quality images with full head coverage and high resolution |
| Face removed raw structural data (MRI, CT)<br>- DISTINCTIVE - | Depends on whether ears and teeth are present; reverse reconstruction can be done (Abramian & Eklund, 2019; Y. Xiao et al., 2022) | ReID: 2.0 SingleOut: 4.0<br>No face blurring; removal should apply. No bone abnormalities should be visible. |
| Skull-stripped raw structural data (MRI, CT)<br>- DISTINCTIVE - | Subject-specific (Takao et al., 2015) | ReID: 2.0 SingleOut: 4.0 |
| Tissue maps (scalp, skull, GM, WM, CSF), VBM, SDM<br>- DISTINCTIVE - | | ReID: 2.0 SingleOut: 4.0 |
| Morphometric maps (thickness, curvature, etc)<br>- DISTINCTIVE - | High single-out rate (Duan et al., 2019; Guillon et al., 2024; Mansour L et al., 2021; Wachinger et al., 2015) | ReID: 2.5 SingleOut: 4.0 |
| Morphometric models (e.g. cortical mesh)<br>- DISTINCTIVE - | | ReID: 2.0 SingleOut: 3.0<br>Consider the image and mesh resolutions. |
| Synthetic images for the same imaging type (e.g. learn from T1w, make T1w)<br>- DISTINCTIVE - | A generative model can memorise images (Dar et al., 2025; Usman Akbar et al., 2025) | ReID: 2.0 SingleOut: 3.0<br>AI model leakage must be assessed (see e.g. Szatkownik et al., 2026) |
| Synthetic images for different imaging types (e.g learn from MRI, make CT; learn from T1w, make T2w)<br>- LOW LIKELIHOOD - | | ReID: 2.0 SingleOut: 2.0<br>Currently not as accurate as generative models from the same modality, but leakage should be checked |
| Table of morphometric (volume, thickness, etc.) or qMRI (T1 time) ROI information | Brains have unique patterns, such that one can single out subjects (Valizadeh et al., | ReID: 1.0 SingleOut: 3.0<br>It is important to consider the number of features (and |
| - DISTINCTIVE - | 2018) | quality) |
| Raw DWI<br>- LOW LIKELIHOOD - | Re-ID has not been demonstrated successfully (Schwarz et al., 2023). | ReID: 2.5 SingleOut: 2.0 |
| Raw dMRI<br>- DISTINCTIVE - | | ReID: 2.5 SingleOut: 4 |
| Derived Diffusion Metrics (e.g. ADC, FA, MD)<br>- LOW LIKELIHOOD - | Track specific features (D. Wang et al., 2023) | ReID: 1.0 SingleOut: 2.5 |
| dMRI connectivity matrices<br>- DISTINCTIVE - | Connectome can single out individuals (Ciarrusta et al., 2022; Mansour L et al., 2021; Yeh et al., 2016) | ReID: 2.0 SingleOut: 4 |

Table 2: structural imaging likelihoods consensus (for participants without visible/noticeable abnormalities)

Functional imaging: BOLD, ASL and fNIRS

Compared with the online ratings, the expert subgroup changed the fNIRS rating, moving single-out likelihoods from 3 to 2, because studies on reproducibility (between sessions and between datasets) point to optode placement as a limiting factor (Novi et al., 2020; Wagner et al., 2025; Yücel et al., 2025) while singling-out studies focused on within single session dataset alone (Zhang & Li, 2024). The consensus left raw fMRI and space-time decomposition (e.g., ICA) undecided but moved the preprocessed data likelihood from 2.5 to 3, having to consider if data are spatially normalised and smoothed (temporally filtered, slice-time corrected, and realigned images similar to raw in terms of likelihoods).

| Brain Imaging type | Evidence | Ratings with common evidence provided N = 20 |
|---|---|---|
| Raw fMRI<br>- DISTINCTIVE - | Face may be visible for very high resolution and high field, but has not been demonstrated successfully (Schwarz et al., 2023) | ReID: 2.0 SingleOut: 3.0<br>Consider the scanner field strength and voxel resolution |
| Raw ASL<br>- LOW LIKELIHOOD - | | ReID: 1.0 SingleOut: 2.0 |
| Preprocessed data<br>- LOW LIKELIHOOD for ASL -<br>- DISTINCTIVE for BOLD - | Brain images retain their main distinguishing information and features (Chen et al., 2013) | ReID: 1.0 SingleOut: 3<br>Consider if images are spatially normalised and smoothed |
| Other perfusion (CE) and Flow imaging (TOF)<br>- LOW LIKELIHOOD - | | ReID: 1.0 SingleOut: 2.0 |
| MRS<br>- LOW LIKELIHOOD - | | ReID: 1.0 SingleOut: 2.0 |
| Task-based activation maps<br>- LOW LIKELIHOOD - | Activation patterns can be stable within a subject (Gorgolewski et al., 2012) but are highly task and motion dependent (Gorgolewski et al., 2013) | ReID: 1.0 SingleOut: 2. |
| Raw or minimally processed time series (fMRI, ASL, fNRIS)<br>- LOW LIKELIHOOD - | A single regional time course is unique enough to single out (Zhang & Li, 2024) | ReID: 2.0 SingleOut: 2.0 |
| Participant-specific ICA and other decomposition outputs - DISTINCTIVE - | Network-specific, like connectivity matrices (N. Wang et al., 2017; X. Wang et al., 2025; Xie et al., 2018) | ReID: 1.5 SingleOut: 3.0 Distinguish participant-specific outputs from group-level component maps. |
| BOLD connectivity matrices<br>- DISTINCTIVE - | Connectomes are unique, even at low resolution (Amico & Goñi, 2018; Byrge & Kennedy, 2019; Finn et al., 2015; Miranda-Dominguez et al., 2014; Van De Ville et al., 2021). The connectivity pattern allows cross-database singling out (Ravindra et al., 2021; Ravindra & Grama, 2019) | ReID: 2.0 SingleOut: 4.0 |
| fNIRS connectivity<br>- DISTINCTIVE - | Sensor-level (Ren et al., 2022; Rodrigues et al., 2019) connectivity is good enough to single out individuals | ReID: 2.0 SingleOut: 3.0 |
| Connectivity summaries (graph-derived metrics, dynamical measures like dwell time, etc)<br>- LOW LIKELIHOOD - | | ReID: 1.0 SingleOut: 2.0 |
| Summary connectivity maps and group matrices<br>- LOW LIKELIHOOD - | | ReID: 1.0 SingleOut: 2.0 |

Table 3: Blood-based functional imaging likelihoods consensus (for participants without visible/noticeable abnormalities)

Positron Emission Tomography

During the discussion, we changed the raw PET imaging likelihood from 2 to 3, given the possibility of face reconstruction. However, the risk of identification is unclear because it depends on many parameters (see Table 4). If in doubt, removing the face, as with structural MRI, is recommended.

| Brain Imaging Information | Evidence | Ratings with common evidence provided N = 20 |
|---|---|---|
| Raw PET images<br>- DISTINCTIVE - | Face can be seen depending on the tracer (Bou Hanna et al., 2024; Schwarz et al., 2022; Selfridge et al., 2023) | ReID: 3.0 SingleOut: 3.0<br>Consider scanner resolution, image reconstruction, tracer, and dose, and test face reconstruction. |
| Parametric maps (e.g., SUV, SUVR, DVR, VT, KI) and kinetic parameters<br>- LOW LIKELIHOOD - | | ReID: 1.0 SingleOut: 2.0 |
| PET connectivity matrices<br>- LOW LIKELIHOOD - | | ReID: 2.0 SingleOut: 2.5 |

Table 4: PET imaging likelihoods consensus (for participants without visible/noticeable abnormalities)

Electrophysiology

No changes to ratings were made during the group discussion. Table 5 presents the classification and considerations.

| Brain Imaging Information | Evidence | Ratings with common evidence provided N = 20 |
|---|---|---|
| Raw and preprocessed signals<br>- DISTINCTIVE -<br><br>Single-trial time courses or features<br>- DISTINCTIVE - | Contains unique features (also in the non-brain signals such as eye blinks, muscle activity), allowing one to single out individuals (da Silva Castanheira et al., 2021; Liu et al., 2024; Meng et al., 2024; Sareen et al., 2021; Wu et al., 2022) | ReID: 2.0 SingleOut: 3.0 |
| Power spectrum<br>- DISTINCTIVE - | Within session highly subject-specific (Castanheira et al., 2025; Demuru & Fraschini, 2020; Jijomon & Vinod, 2018; Stergiadis et al., 2022; Suppiah & Vinod, 2018; Tatar, 2023; Thomas & Vinod, 2017) and matching across independently acquired sessions (da Silva Castanheira et al., 2021; DelPozo-Banos et al., 2015) | ReID: 2.5 SingleOut: 4.0 |
| ERP/EFP<br>- DISTINCTIVE - | ERPs are highly reliable within subjects (Cassidy et al., 2012; Gaspar et al., 2011) but moderately stable over time (Weinberg & Hajcak, 2011) | ReID: 1.0 SingleOut: 3.0<br>Depends highly on the stimulus type used |
| MEEG connectivity matrices<br>- DISTINCTIVE - | connectome can single out individuals (da Silva Castanheira et al., 2021; Fraschini et al., 2015; Kong et al., 2019; Rocca et al., 2014) | ReID: 2.0 SingleOut: 4.0 |

Table 5: MEG-EEG likelihoods consensus (for participants without visible/noticeable abnormalities)

# Discussion

The expert consensus identified 34 brain imaging items commonly used and shared, with only 1 classified as identifiable: raw unaltered structural MRI/CT data. Among the other items, 20 were classified as distinctive and 13 as low likelihood of re-identification. Among the 20 distinctive items, 10 have singled out ratings of 3 (i.e. borderlines) with 8 cases given considerations, meaning that at minimum 38% and up to 67% of neuroimaging data types presented here can be shared with minimal restriction, in regard to their intrinsic re-identification likelihood.

Because of the limited availability of publicly accessible, identity-linked brain reference datasets, we concluded that individuals without noticeable skull or brain abnormalities face very low identification risk via linkage. Unaltered raw structural MRI/CT and occasionally PET scans remain the exception, because faces can be reconstructed and matched with public pictures (e.g., from social media). Aside from faces, our view is that, when sharing brain imaging data (in particular derivatives, see Tables 2, 3, 4, 5), the risk is not, in theory, higher than using behavioural profiling (i.e. the passive, unintentional trail of data people leave behind as they interact with digital systems, such as clickstreams, scroll paths, typing rhythms, timestamps, and device metadata) which are used daily by digital companies to single out individuals to target advertising. Our results complement the Medical Image De-Identification (MIDI) Task Group recommendations (Clunie et al., 2025) which address technical procedures for reducing re-identification risk before the public release of medical images (including ‘source’ data such as DICOM files). While MIDI considers both image content and associated identifying information, our consensus focuses specifically on the identification-relevant properties of neuroimaging measurements and their derivatives, assuming that directly identifying metadata and annotations have been removed, which typically is the case during conversion from, e.g. DICOM to NIfTI. Such metadata removal is essential and has been shown to be effective in preventing re-identification (Catalán Flores et al., 2026). Both, our assessment and the MIDI task force, distinguish the potential to match individual brain images across datasets from the practical ability to associate those images with a named individual. In particular, the lack of widely available, identity-linked reference databases of brain anatomy or function (exceptions are known brains, e.g. Holmes et al., 1998; Poldrack et al., 2015) limits opportunities to re-identify individuals from brain-derived features, despite their demonstrated individual specificity.

The likelihood of identification from neuroimaging data depends not only on the imaging representation but also on acquisition characteristics, the availability of external reference information, and the capabilities of potential attackers. For structural images, facial reconstruction and recognition depend on anatomical coverage, spatial resolution and acquisition parameters. Image modifications such as defacing or skull stripping can reduce facial identification risk but can also affect downstream analyses (Gao et al., 2023) and do not eliminate individual-specific information contained within brain anatomy. Consequently, decisions about image modification and sharing should consider both residual identification likelihood and preservation of scientific utility. Defacing is, however, an effective technique, and empirical work demonstrated its effectiveness (Steeg et al., 2024). We, therefore, propose that the present consensus ratings should be interpreted as evidence to inform context-specific assessments, rather than as fixed properties of imaging data. Such assessments may require periodic revision as imaging technology, identification methods and the availability of reference datasets evolve. From an imaging standpoint, it is more straightforward for someone knowledgeable to identify participants, since uniqueness can increase drastically in such cases. Federated analysis and learning has been proposed as an alternative to enable collaborative research strategies, retaining participant-level data locally (Pati et al., 2022; Plis et al., 2016) although shared parameters and outputs still require assessment for potential information leakage (Gupta et al., 2021). Yet another approach to facilitate brain research is to share realistic synthetic images instead of real images. Synthetic images are currently a legal gray zone, as a synthetic image generated from noise does not belong to a specific person and by definition is not personal data. However, a synthetic image can be very similar to a real image used for training the generative model (Usman Akbar et al., 2025, Dar et al., 2025), and can then be seen as personal data (rated as distinctive or low likelihood depending on use case, see Table 2).

During the nominal group and DELPHI methodologies, different interpretations of risk (absolute vs relative), identifiability, and single out emerged among experts, and even within experts depending on the item rated (see items generation results). According to the Cambridge Dictionary, the term "identification" refers to the ability to recognise, name, or prove who someone is. One issue when reviewing the literature is that recognising and naming implies retrieving a full identity, such as matching someone’s face from an MRI to photos on social media. This is what we call re-identification. By contrast, proving who someone is does not always equate to recognition or naming. Brain imaging can show individual-specific patterns stable enough to single people out across sessions and datasets. For instance, someone's connectome has unique features that could let us infer that a person from one study is the same person scanned in another (Amico & Goñi, 2018; Ravindra & Grama, 2019). Yet, many of the articles referenced in tables 2, 3, 4 and 5 talk about re-identification or fingerprinting (see glossary), while in reality they simply demonstrate uniqueness of brain imaging data. This is not trivial because unlike imaging experts, data protection officers and legal representatives will likely interpret those results as demonstrating identifiable data, rather than distinctive. We urge the imaging community to use more specific vocabulary when presenting results. While using ‘identification’ makes articles’ titles more striking, it does harm long term data sharing and the scientific endeavour.

To use our assessments in deciding on what policy to use when sharing data, it is necessary to consider that we focused (1) purely on imaging data types; and (2) only brains without visually apparent abnormalities (therefore including patients with common neuropsychiatric disorders). Imaging datasets rarely contain only imaging data, and metadata, particularly demographics, must also be considered (Kibsgaard et al., 2025). Tools such as metaprivBIDS (Kibsgaard & Pernet, 2026) can identify at-risk variables and participants and reduce their associated re-identification risk, for example, through binning and noise masking. This issue extends beyond inference from background or linkage attacks on demographic data because brain imaging data (MRI and MEEG) can be used to infer age and sex (Darvishi-Bayazi et al., 2025; Dörfel et al., 2026; Ebel et al., 2023; Vandenbosch et al., 2019). When sharing imaging data from participants with noticeable skull or brain abnormalities (e.g. tumours, strokes), the ratings given in this paper do not apply directly and an evaluation should consider higher inferential risks. Our ratings may, however, apply to other neurological conditions without distinguishing individual patterns such as multiple sclerosis, various dementia diseases, or epilepsy cases, providing associated demographics and medical information are curated to prevent inference (see e.g. re-identification risks when too much associated data are given with the UK-BioBank leak, Devlin et al., 2026)). Such usage is well illustrated by the iEEG rating (see figure 1, row 9) which are data most often collected in epilepsy patients. While not included in the final consensus (since focusing on participants without visible/noticeable abnormalities), it shows a median expert re-identification rating of 2 and of single-out of 3.5 which would classify this data type as distinctive and thus possibly shareable with limited restrictions depending on the patients' characteristics. Finally, for research that could generate important clinical information about individual participants, arrangements should allow a party (e.g., a database manager) to hold a key to re-identify data sources if needed, under stringent privacy protections (McCarty et al., 2008) along with all necessary organisational measures (Eke et al., 2021).

# Acknowledgments

CP and MG were supported by the Novo Nordisk Foundation (OpenNeuroPET: grant agreement NN20OC0063277). CP and R.O. were supported by the European Union's Horizon Research and Innovation programme (SIESTA: grant agreement 101131957). P.R. acknowledges support from the EU4Health Programme through SCAILED (101314097),EU Horizon Europe Programme through BRIDGE (101219311), EBRAINS2.0 (101147319), Virtual Brain Twin (101137289), EBRAINS-PREP(101079717), AISN (101057655), EBRAIN-Health (101058516), and EIC grant PHRASE (101058240); the Digital Europe Programme through TEF-Health (101100700), SHAIPED (101195135), and CoordinaTEF (101168074); the German Research Foundation (DFG) through SFB 1436 (Project ID 425899996), SFB 1315 (Project ID 327654276), SFB 936 (Project ID 178316478), SPPComputational Connectomics (RI 2073/6-1, RI 2073/10-2, and RI 2073/9-1), DFG Clinical Research Group BECAUSE-Y (504745852), the DFG German-Spanish project BrainsAge (RI 2073/14-1), and DFG Project ID 424778381 – TRR 295 Retune; ERA PerMed JTC2021; the German Federal Ministry for Health through PatternCog (ZMI5-2522FSB904); Berlin University Alliance OpenMake; and the Berlin Institute of Health and Foundation Charité.

# Glossary

**Anonymisation:** Processing personal data so the individuals concerned are no longer identifiable under the applicable legal standard, taking into account the relevant context and available identification methods.

| Framework | Terminology and threshold |
|---|---|
| EU/GDPR and UK GDPR | Anonymisation renders individuals no longer identifiable, considering the means reasonably likely to be used. The resulting anonymous information falls outside the respective GDPR's scope. |
| US/HIPAA | Uses de-identification, rather than anonymisation, for information meeting §164.514 through Safe Harbor or Expert Determination. The information is then no longer protected health information under HIPAA. This is a distinct legal test, not automatically equivalent to GDPR anonymity. |
| Australia | The Privacy Act uses de-identification: individuals must no longer be identifiable or reasonably identifiable. OAIC assesses this contextually; removing direct identifiers alone is insufficient. |
| Québec | Expressly defines anonymised information: in the circumstances, it must remain reasonably foreseeable that individuals can irreversibly no longer be identified directly or indirectly. Prescribed practices and regulatory requirements also apply. |
| Singapore | PDPC guidance defines anonymisation as converting personal data into information that cannot identify an individual, using a risk-based combination of techniques and safeguards. |

**Brain fingerprinting:** expression used in brain imaging studies to indicate the uniqueness of brain patterns. Unlike actual fingerprinting, it cannot lead to identification because there are no databases of 'brain-prints'.

**De-identification:** An umbrella term for processes that reduce individuals' identifiability in data, for example by removing or modifying identifiers and applying additional safeguards. The term does not have a uniform legal meaning: under HIPAA and Australia's Privacy Act, successful de-identification must meet a specified identification threshold, whereas other frameworks use the term for processing that may leave individuals identifiable. Unless a particular legal framework is specified, "de-identified" does not, in this manuscript, imply that data are anonymous or outside applicable privacy law.

| Framework | Meaning of de-identification | Consequence |
|---|---|---|
| US—HIPAA, 45 CFR §164.514 | Health information must neither identify an individual nor provide a reasonable basis | Successfully de-identified information is no longer |
| | for identification. This requires Expert Determination or Safe Harbor: removal of specified identifiers plus absence of actual knowledge that the remaining information could identify someone. | protected health information under HIPAA. Removing names alone is insufficient. |
| Australia—Privacy Act 1988, §6(1) | Information is de-identified when it is no longer about an identifiable individual or an individual who is reasonably identifiable. OAIC guidance considers the data, available additional information, recipients and access environment. | Appropriately de-identified information is no longer personal information under that Act. This is an outcome threshold, rather than simply removal of direct identifiers. |
| EU—GDPR | De-identification is not a defined GDPR category. The relevant distinctions are personal data, pseudonymisation and anonymous information. | Calling data "de-identified" does not establish their legal status; the applicable identifiability assessment remains necessary. |
| UK—DPA 2018, §171 | For the provisions concerning unlawful re-identification, data are de-identified when they cannot be attributed to a specific person without further information. ICO discourages using the term as a general synonym for anonymous or pseudonymous data. | The term can describe data that remain personal data. |
| Canada—federal government guidance | A process of modifying identifiers and applying controls to reduce identifiability appropriately for the context. The guidance explicitly recognises residual re-identification risk and distinguishes de-identified from anonymised information. | De-identification does not automatically remove information from the Privacy Act's scope; sufficiently extensive processing may do so following assessment. This is federal administrative guidance, not a definition covering every Canadian privacy regime. |
| Singapore—PDPC guidance | De-identification means removing identifiers that directly identify someone. It is described as only the first step towards anonymisation. | De-identified data can remain personal data, including when identification is possible through combination with other information. |

**Identifiability** - Identifiability is the possibility of identifying a data subject. Identifiability is not only direct identification of an individual but can also include indirect identification, such as through singling out.

**Personal data** - Personal information is any data that pertains to an identified or identifiable person.

**Pseudonymisation:** Under GDPR Article 4(5), processing personal data so that attribution to a specific individual requires additional information, which is kept separately and protected by technical and organisational measures. Pseudonymisation does not itself determine anonymity: the resulting data remain personal for entities able to identify individuals using means reasonably likely to be used, but may be anonymous for other recipients, depending on the processing context.

**Re-identification:** the process of attributing an identity to a record that has been anonymized/pseudonymized or de-identified.

**Singling-out:** In the EU, "record isolation" denotes distinguishing an individual record within a dataset, whereas "singling out" denotes distinguishing the corresponding individual in the relevant context using means reasonably likely to be used. Related distinctions appear in other frameworks through concepts such as distinguishability, linkage and reasonable identifiability.

| Framework | Terms used | Definitions and relationship to the EDPB distinction |
|---|---|---|
| EU—EDPB Guidelines 02/2026 (version 1.0) | Record isolation; singling out | Record isolation: a unique combination of attribute values distinguishes an individual record within the given dataset. Singling out: an isolated record can be matched to the corresponding individual's attributes, distinguishing that individual from others in the relevant context using means reasonably likely to be used. Record isolation alone does not establish identifiability; discovering the person's name is not required. These are distinctions used in EDPB guidance, rather than separately defined terms in GDPR Article 4. |
| UK—ICO guidance | Singling out; distinguishable or isolated records; linkability | Very close conceptually. ICO explicitly states that distinguishing one record from others is insufficient by itself to make the person identifiable. It also considers whether information enables someone to discover identity or act specifically on the person without knowing their name. It does not organise these concepts into the same formally named pair. |
| US—HIPAA/HHS guidance | Distinguishability; uniqueness; linkage; identification risk | HHS separates distinguishable data from successful identification. Its Expert Determination guidance considers whether distinguishing features also occur in accessible identified sources and whether a matching mechanism exists. "Record isolation" and "singling out" are not the named legal tests. |
| Australia—Privacy Act/OAIC guidance | Unique or rare characteristics; | The distinction is incorporated into a contextual assessment rather than separately named. |
| | reasonably identifiable individuals | Uniqueness contributes to risk, while reasonable identifiability also depends on available information, access, practicality and the likelihood of identification. The guidance reviewed does not define “record isolation” and “singling out” as a formal pair. |
| Québec—2024 anonymisation regulation | Individualization; correlation; inference | The regulation defines its individualization criterion as the inability to isolate or distinguish a person within a dataset. It then requires a broader re-identification assessment involving reasonably available information, resources, expertise and circumstances. It does not explicitly separate record isolation from person-level singling out in the EDPB’s manner. |
| Canada—federal terminology guidance | Individualization; singling out as an alternative designation | The government glossary describes isolating a person within a dataset without using other datasets. This usage is closer to the older, broader meaning of singling out; it should not automatically be equated with the EDPB’s refined distinction. This is terminology guidance, rather than a statutory definition. |

**Special categories of personal data** - Personal data has a number of categories that are deemed to be special; these are sometimes also referred to as sensitive data.

# Annex 1

Each expert prepared a table listing different imaging types and derived data, along with risk rating, opinion, evidence, and expertise level. This nominal group approach, which involved listing data types alone, was combined to produce semantically homogeneous lists shared for the DELPHI rating. The following email was sent:

*Title*: expert consensus paper on imaging data privacy

Dear friend and colleague,

We have been in touch before, or a colleague recommended you to join this work. If you are (still) willing to participate, we are interested in getting your expert opinion on the likelihood of identification for neuroimaging data. The goal is to write a recommendation paper that we, brain imagers, can use to discuss with data protection officers and lawyers (who need such information) when sharing data.

We are combining the nominal group approach of listing with the DELPHI rating to reach consensus. The work will be carried out in 3 rounds. The 1st round is the most time-consuming (see below). After that, similar work is to be carried out for a small subset for which you have the most expertise. The 3rd round will be an online group discussion in your area of expertise.

Expected timeline:

- Round 1 from now until Christmas (i.e at your convenience, find 1 hour or so to make a table and send it to me)
- Round 2 in March/April, time for me to assemble all data and re-distribute as appropriate
- Round 3 sometime before OHBM in May/June

<u>For now, I am asking you to create de novo a list with ratings:</u>

(A) List all the data types you can think of that can be shared, covering MRI, PET, EEG, MEG and fNIRS. Assume we share curated data with no names, addresses, etc (image/electrophysio data metadata are clean). In your list, include converted data (no k-space, no dicom, *we are talking nifti, edf, etc*) and any derived data (data computed from converted data).

(B) For each data type, rate the risk of identification from 1 to 5, with 5 being high likelihood (data can be seen as personal even after pseudonymization, they contain directly identifiable features, or features that can easily lead to reidentification either because of a low technical barrier or easy linkage with public sources), 3 being medium risk (data have been pseudonymized, they contain unique features and this could allows to single out individuals but such feature either requires expertise to extract or the linkage risk is deemed low) and 1 being no risk (data are fully anonymous and have no uniquely identifiable features).

(C) Consider the following four markers when rating likelihood: direct identification, identification via background attack (ie what if you know someone took part in a study, can you identify just by looking at the data), identification via linkage attack (can someone link the imaging data with another source to identify someone) and finally, what is the technical know-how to identify (if feasible)

(D) Please try to include any references/evidence you have that informed your decision (doi, pubmed id, whatever you have, I'll deal with formatting/accessing the information)

(E) Finally, indicate your own expertise in the data type from 1 (expert) to 5 (you just know about it).

Here is a small example from my own list (small on purpose, as I am asking you what derived data you know about)

| Data type | Risk | Risk type and evidence<br>Direct - background - linkage -technical know-how | My expertise |
|---|---|---|---|
| Raw Structural MRI | 5 | See references in the open brain consent paper - face matching is the biggest risk via background attack or linkage; from background attack, it is easy to reconstruct a face with little knowledge - linkage to photograph still requires heavy computing (but all pen source tools reach 60% accuracy https://www.sciencedirect.com/science/article/pii/S1053811925004793?via%3Dihub#sec0009) | 2 |
| Table of volumetric information per ROI | 1 | There is close to zero likelihood IMO since no one knows each other's volumes, thicknesses, etc .. and there is no database to link to | 2 |
| Defaced Structural MRI | 2 | There are still plenty of unique features, risk will also depend on the population, and it requires more expertise + medium linkage risk with other databases like dental records | 2 |
| Preprocessed fMRI (realigned, spatially normalised) | 2 | While time courses and connectivity from that can be unique, being in MNI space removed the anatomical marker, linkage inexistent (assuming the reverse normalisation field is not shared) | 1 |
| Connectivity matrices based on DWI | 1 | | 3 |
| Raw PET data | 4 | (depends on the tracer, with some tracers the signal is high in the face, which can be reconstructed like MRI - there is 1 or 2 papers on this) | 3 |
| Evoked Responses from EEG or MEG | 2 | (see references in the open brain consent paper - these are unique for sure, but there are no databases to link to) | 1 |

Best whishes
Cyril

# Annex 2

The imaging types followed broad distinctions: MRI (structural, diffusion, and dynamic imaging), PET, Electrophysiology (MEG and EEG together), and connectomics. This distinction was used for the DELPHI rating ordering items within those classes. The table below reports all items generated in round one with raw ratings for full disclosure, but the DELPHI rating included fewer items because some were grouped (see results).

| Brain Imaging Information<br>(not considering the uniqueness of patients' brains) | Individual ratings, median and IQR. |
|---|---|
| Raw structural MRI | Ratings: 5, 5, 5, 5, 4, 5, 5, 5, 5, 5, 5, 5, 2, 5, 5, 5, 5, 5, 5, 5<br>N= 20, M= 5, IQR= 0 |
| Raw structural CT | Ratings: 5, 5, 3<br>N= 3, M= 5, IQR= 2 |
| Face blurred raw structural data (MRI, CT) | Ratings: 4, 4, 4, 4<br>N= 4, M= 4, IQR= 0 |
| Face removed raw structural data (MRI, CT) | Ratings: 2, 4, 3, 3, 2, 3, 3, 2, 4, 2, 2, 2, 1, 2, 3, 3, 3, 2, 3, 3<br>N= 20, M= 3, IQR= 1 |
| Skull-stripped raw structural data (MRI, CT) | Ratings: 1, 4, 3, 2, 3<br>N= 5, M= 3, IQR= 2 |
| Spatially normalised structural data (MRI, CT) | Rating: 4 |
| Tissue maps (scalp, skull, GM, WM, CSF), VBM, SDM | Ratings: 3, 2, 1, 2<br>N= 4, M= 2, IQR= 1.5 |
| Lesion mask | Rating: 1 |
| Morphometric models (e.g. cortical mesh) and maps (thickness, curvature, etc) | Ratings: 2, 4, 2, 3<br>N= 4, M= 2.5, IQR= 1 |
| Synthetic structural MRI and CT for the same imaging type (e.g. learn from T1w to make T1w) | Ratings: 2, 3<br>N= 2, M= 2.5, IQR= 1 |
| Synthetic structural MRI and CT for different imaging types (e.g learn from MRI, make CT; learn from T1w, make T2w) | Ratings: 4, 1<br>N= 2, M= 2.5, IQR= 3 |
| Table of morphometric (volume, thickness, etc.) or qMRI (T1 time) information (ROI based) | Ratings: 1,1,1, 3, 2, 1, 2, 2, 2, 1, 1, 1,1<br>N= 13, M= 1, IQR= 1 |
| Raw DWI | Ratings: 3, 4, 4, 2, 2, 2, 1, 3<br>N= 8, M= 2.5, IQR= 1 |
| Scanner-derived DWI (e.g. FA, MD) | Ratings: 4, 1, 2, 3<br>N= 4, M= 2.5, IQR= 2 |
| Track profiles | Rating: 2 |
| Raw PET images | Ratings: 4, 3, 4, 3, 4, 4, 1, 2, 3<br>N= 9, M= 3, IQR= 1.5 |
| Skull-stripped PET images | Ratings: 3 |
| Regional/Voxex/Vertices Time–Activity Curves | Ratings: 1, 2, 1<br>N= 3, M= 1, IQR= 1 |
| Parametric maps (e.g., SUV, SUVR, DVR, VT, KI) and kinetic parameters | Ratings: 1, 1, 2<br>N= 3, M= 1, IQR= 1 |
| Blood data (e.g. tracer concentrations over time) | Rating: 1 |
| Raw fMRI | Ratings: 3, 2, 3, 2, 1, 1, 3, 3<br>N= 8, M= 2.5, IQR= 1 |
| Raw ASL | Rating: 3 |
| K-space fMRI / ASL | Rating: 1 |
| Other perfusion (CE) | Rating: 4 |
| Flow imaging (TOF) | Rating: 3 |
| MRS | Rating: 2 |
| Preprocessed data, subject space | Ratings: 3, 2, 2, 3, 2, 3, 2, 2, 3<br>N= 9, M= 2, IQR= 1 |
| Spatially normalised (raw or preprocessed) | Ratings: 2, 1, 2, 3, 1, 2, 1, 1, 1, 1, 2, 2<br>N= 12, M= 1.5, IQR= 1 |
| ROI time courses | Ratings: 2, 1, 1<br>N= 3, M= 1, IQR= 1 |
| ICA and other decompositions (time courses and topographies) | Ratings: 2, 2<br>N= 2, M= 2, IQR= 0 |
| Head motion parameters | Ratings: 1, 1<br>N= 2, M= 1, IQR= 0 |
| Motion tracking data cloud | Rating: 5 |
| Task-based activation maps | Ratings: 1, 1, 4, 1, 5, 1, 1, 1, 3<br>N= 9, M= 1, IQR= 2.5 |
| Raw and preprocessed signals | Ratings: 2, 2, 3, 3, 3, 3, 2, 3, 2<br>N= 9, M= 3, IQR= 1 |
| iEEG | Ratings: 3, 3<br>N= 2, M= 3, IQR= 0 |
| Single-trial time courses or features | Ratings: 3, 2<br>N= 2, M= 2.5, IQR= 1 |
| Power spectrum | Ratings: 3, 2, 3, 2<br>N= 4, M= 2.5, IQR= 1 |
| ERP/EFP | Ratings: 1, 1, 3, 2, 3, 1, 2<br>N= 7, M= 2, IQR= 2 |
| ICA and other decompositions (time courses and topographies) | Rating: 1 |
| MEEG head models | Ratings: 2, 3<br>N= 2, M= 2.5, IQR= 1 |
| E-field, M-field magnitude/direction maps from MRI head models | Rating: 2 |
| Source reconstructed MEEG | Rating: 3, 3, 2<br>N= 3, M= 3, IQR= 1 |